\documentclass[aip,reprint,graphicx]{revtex4-1}

\usepackage{graphicx}

\begin{document}

%\preprint{}

\title{Field-induced continuous rotation of the polarization in multiferroic Mn$_{0.95}$Co$_{0.05}$WO$_{4}$}



%\thanks{}

\author{K.-C. Liang}
\email{kliang@mail.uh.edu}
\author{R. P. Chaudhury}
\altaffiliation[Current address: ]{Intel Corporation, 2501 N.W. 229th St., Hillsboro, OR 97124, USA}
\author{Y. Q. Wang}
\author{Y. Y. Sun}
\author{B. Lorenz}
\author{C. W. Chu}
\altaffiliation[Also at: ]{Lawrence Berkeley National Laboratory, 1 Cyclotron Road, Berkeley, California 94720, USA}
\affiliation{ $TCSUH$ and Department of Physics, University of Houston, Houston, Texas 77204-5002, USA}


\date{\today}

\begin{abstract}
 We report the observation a continuous rotation of the polarization in Mn$_{0.95}$Co$_{0.05}$WO$_{4}$ under magnetic field. At zero field, this compound shows a transition into the spiral magnetic and ferroelectric phase at 12.2K, which is the ground state, with the polarization oriented along the $b$-axis. Increasing $b$-axis magnetic fields rotate the ferroelectric polarization continuously toward the $a$-axis, indicating a rotation of the spin spiral plane. This rotation extends over a large field and temperature range. At a constant magnetic field of 3 Tesla, the polarization also rotates from the $a$-axis at the ferroelectric transition toward the $b$-axis upon decreasing temperature.
\end{abstract}


\maketitle %\maketitle must follow title, authors, abstract and \pacs


%\section{Introduction}
Multiferroics are materials that exhibit the strong coupling and coexistence of ferroelectric (FE) and magnetic orders and, therefore, could bear the potential to have a large magnetoelectric (ME) effect. The ME effect allows for the control of electric polarization (magnetization) by magnetic (electric) fields.\cite{Fiebig:05} Recent studies of such cross correlation in frustrated magnets, such as flipping or drastic changes of the FE polarization (P) under magnetic fields, have attracted significant scientific interest on multiferroic compounds and its applications.\cite{Cheong:07} It has been shown that the antisymmetric exchange interaction, i.e., the well known Dzyaloshinskii-Moriya interaction plays the key role when ferroelectricity is induced by inversion symmetry breaking spiral (noncollinear) magnetic orders\cite{Mostovoy:06,Katsura:05} in systems such as TbMnO$_{3}$, Ni$_{3}$V$_{2}$O$_{8}$, MnWO$_{4}$, etc.

The spiral spin phase (AF2) in MnWO$_{4}$ forms between 7.8 and 12.6 K, and it is sandwiched between an incommensurate (ICM) sinusoidal AF3 phase (12.6 K$<$T$<$13.5 K) and a commensurate (CM) AF1 phase (T$<$7.8 K). The AF2 and AF3 phases are both ICM with similar wave vectors $\roarrow q_{2,3}$ =(-0.214, 1/2, 0.457). The AF1 phase is collinear [q$_{1}$ =($\pm $1/4, 1/2, 1/2)] with a characteristic frustrated spin structure $\uparrow\uparrow\downarrow\downarrow$.\cite{Lautenschlager:93} Only the AF2 phase is ferroelectric, which can be explained by the spatial inversion symmetry breaking spiral spin structure and existing strong spin-lattice interactions.\cite{Taniguchi:06}

The complex magnetic orders including the spin spiral in multiferroic systems have their origin in competing magnetic exchange interactions leading to frustration and a remarkable sensitivity of the frustrated states with respect to small perturbations such as magnetic fields, pressure, or substitutions. In the earlier report, a small amount Co$^{+2}$ substitution was observed to stabilize the AF2 phase as the ground state and to suppress the commensurate AF1 phase based on the magnetic and neutron scattering experiments of polycrystalline samples.\cite{Song:09} The neutron scattering data further suggested the rotation of the FE polarization from the $b$-axis toward the $a$-axis at higher Co content. A sizable $a$-axis polarization was indeed found in a single crystal of Mn$_{0.9}$Co$_{0.1}$WO$_{4}$ very recently.\cite{Song:10} At slightly higher doping (15 \% Co), however, the only component of the FE polarization was aligned with the $b$-axis.\cite{Chaudhury:10} These seemingly conflicting results indicate an extreme complexity of the multiferroic phase diagram of Mn$_{1-x}$Co$_{x}$WO$_{4}$ and warrant further exploration of the effect of Co substitution on the multiferroic properties of Mn$_{1-x}$Co$_{x}$WO$_{4}$ single crystals.

%\section{Experimental}
Large crystals of Mn$_{0.95}$Co$_{0.05}$WO$_{4}$ have been grown using the floating zone method. The homogeneity of the crystals and their Co content were verified by energy-dispersive X-ray (EDX). EDX results clearly confirm the Co concentration of the sample to be 0.05. A small oriented piece of sample of area 20 mm$^2$ and thickness 1 mm was prepared for measurements. The sample was mounted on a home-made probe inserted into the Physical Property Measurement System (quantum  Design) for precise control temperature and magnetic fields. The pyroelectric current was measured by the electrometer K6517 (Keithley) at a temperature rate of 1 K/min upon warming. An electric field of 1.5 kV/cm (poling voltage) was applied to align the FE domains during cooling. The polarization is than calculated by integrating the current from high temperature (in the paraaelectric state) to low temperatures.

%\section{Results and discussion}
At zero magnetic field, Mn$_{0.95}$Co$_{0.05}$WO$_{4}$ shows the two transitions into the AF3 and AF2 phases, however, the commensurate AF1 phase which forms the ground state in MnWO$_{4}$ is missing. The ferroelectric (spiral) AF2 phase extends from $T_{2}$=12.2K to the lowest temperature and becomes the ground state in Mn$_{0.95}$Co$_{0.05}$WO$_{4}$. In the AF2 phase, the spins form a helical structure, the helical plane is defined by the $b$-axis and the spin-easy axis in the $a-c$ plane, the latter forming an angle of 15.8$^o$ with respect to the $a$-axis.\cite{Song:10} The multiferroic properties of Mn$_{0.95}$Co$_{0.05}$WO$_{4}$ have been investigated in magnetic fields applied along $b$-axis.

\begin{figure}
\includegraphics[width = 3in]{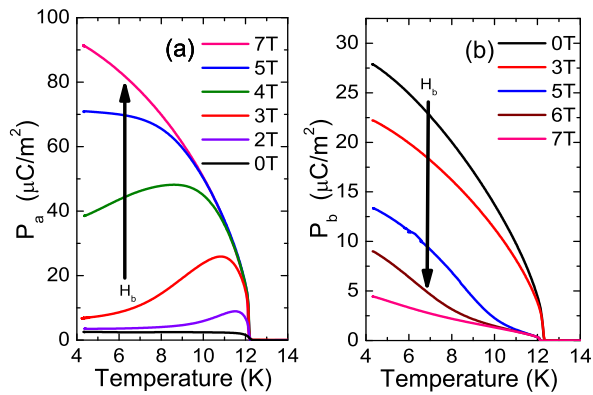}
\caption{(Color online) Ferroelectric polarization of Mn$_{0.95}$Co$_{0.05}$WO$_{4}$ in magnetic fields oriented along the $b$-axis. (a) $a$-axis component (b) $b$-axis component. The values of the magnetic field in Tesla are listed next to the curves.}
\end{figure}

Measurement of the $b$-axis polarization as a function of temperature at different magnetic fields is shown in Fig. 1b. At zero magnetic field, the major component of the FE polarization is still aligned with the $b$-axis. A small $a$-axis component of $P_{a}$=2.3 $\mu C/m^2$ was also measured. The application of $b$-axis magnetic field shows a drastic change of the ferroelectric properties. With increasing $H_{b}$, the polarization along the $b$-axis decreases continuously and $P_{a}$ increases quickly. The $a$-axis polarization shows an interesting temperature dependence in magnetic fields. At lower fields, $P_{a}$(T) increases sharply at $T_{C}$ and passes through a maximum before decreasing to a smaller value at low T. Only above 5 Tesla, $P_{a}$ increases gradually and it reaches its maximum of 90 $\mu C/m^2$ at 7 Tesla and 4.2 K. It is worth noting that this value is nearly identical to the zero-field $P_{a}$ of the higher doped sample with x=0.1 \cite{Song:10}. From the results shown in Fig. 1, it can be concluded that the magnetic field rotates the macroscopic ferroelectric polarization of Mn$_{0.95}$Co$_{0.05}$WO$_{4}$ continuously from the $b$-axis toward the $a$-axis with a simultaneous increase of its magnitude from about 30 $\mu C/m^2$ to 90 $\mu C/m^2$.

The origin of the change of the components of the FE polarization along $a$- and $b$-axes in magnetic fields needs a more detailed consideration. There arises the question whether P$_{a}$ and P$_{b}$ are components of the same polarization vector of a single magnetic phase or they originate from a phase separation into two helical phases with different orientations of the helical plane. Preliminary neutron data for Mn$_{0.95}$Co$_{0.05}$WO$_{4}$ in high magnetic fields have not shown evidence for a phase separation into two magnetic structures.\cite{Ye:11} It is therefore justified to attribute the two components $P_{a}$ and $P_{b}$ (in Fig. 1) to the $a$- and $b$-components of one vector $\roarrow P$. This leads to the conclusion that the $b$-axis magnetic field rotates the spin helical plane by 90 $^\circ$ into the $a-c$ plane at the maximum field.

\begin{figure}
\includegraphics[width = 2.78in]{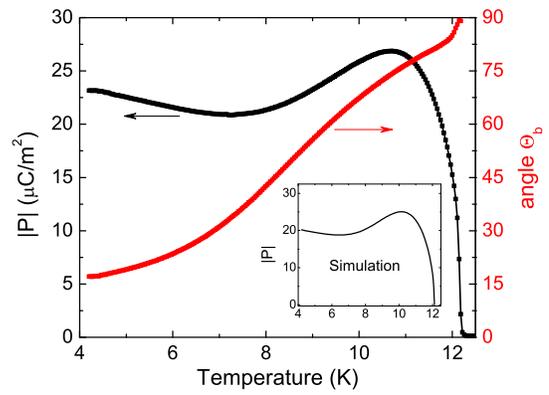}
\caption{(Color online) Magnitude and orientation of the polarization vector at an external field of 3 Tesla. $\Theta_{b}$ is the angle of the polarization vector with the $b$-axis. The inset shows the polarization magnitude calculated from equation 4.}
\end{figure}

The magnitude of this vector, $\left|P\right|$, and its angle with the $b$-axis, $\Theta_{b}$, can then be calculated as a function of temperature and magnetic field using the experimental data for $P_{a}$ and $P_{b}$. The results of $\left|P\right|$ and $\Theta_{b}$ measured at a constant field of 3 Tesla as a function of temperature are shown in Fig. 2. The magnitude of the polarization rises quickly below $T_{C}$ and displays a pronounced maximum at about 11 K. At lower temperature $\left|P\right|$ tends to saturate at a value of 23 $\mu C/m^2$ after passing through a shallow minimum. This unusual behavior of $\left|P(T)\right|$ is accompanied by a smooth change of $\Theta_{b}(T)$ from 90$^o$ at $T_{C}$ to about 15$^o$ at 4 K. The ferroelectric polarization is first aligned with the $a$-axis at $T_{C}$ but it rotates toward the $b$-axis with decreasing temperature.

The complex temperature dependencies of $\left|P(T)\right|$ and $\Theta_{b}(T)$ (Fig. 2) have to be explained by the rotation of the helical plane defining the magnetic order of the Mn spins. The rotation of the spin helix about the easy axis of magnetization is quantitatively described by the angle $\beta$ of the helical plane with the $b$-axis. At the field of 3 Tesla, just below the ferroelectric transition temperature, the spin helix apparently lies in the $a-c$ plane ($\beta(T_{C}) = 90^o$) and the polarization is aligned with the $a$-axis. With decreasing temperature, however, $\beta$ decreases and the helical plane rotates toward the $b$-axis. At the same time, the magnitude of the polarization, $\left|P(T)\right|$, increases as expected, for example, form the Landau theory of phase transitions. The unusual temperature dependence of $\left|P(T)\right|$ at 3 Tesla (Fig. 2) with a maximum at 11 K must be a combined effect of both, the increase of $\left|P(T)\right|$ with decreasing $T$ and the rotation of the polarization vector from the $a$-axis to the $b$-axis.

To elucidate this phenomenon in more detail we have to describe the polarization change with the rotation of the spin-helical plane. Sagayama et al.\cite{Sagayama:08} have derived a simple formula for the $b$-axis FE polarization of MnWO$_{4}$ employing the spin-current model\cite{Katsura:05:1} and adding all contributions along the chains of nearest neighbor Mn ions stretching along the $c$-axis.
\begin{equation}
\roarrow P = {\frac{2A}{V_{cell}}}m_{b}m_{easy}e_{c}sin(\pi q_{z})sin(\alpha)\roarrow b
\end{equation}
$m_{b}$ and $m_{easy}$ are the two orthogonal components of the magnetic moment vector along the $b$-axis and the easy axis, respectively, defining the ellipticity of the helical spin order and $q_z$ is the magnetic modulation along the chains. $A$, $V_{cell}$, and $e_{c}$ are constants, in part related to the structural details. In MnWO$_{4}$ the easy axis lies in the $a-c$ plane at an angle $\alpha$=34$^o$ with the $a$-axis. This formula can be generalized to include the rotation of the helical plane by an angle $\beta$ away from the $b$-axis,
\begin{equation}
\roarrow P = {\frac{2A}{V_{cell}}}m_{b}m_{easy}e_{c}sin(\pi q_{z})(sin(\beta)\roarrow a+cos(\beta)sin(\alpha)\roarrow b)
\end{equation}
$\roarrow a$ and $\roarrow b$ are the unit vectors along the $a$- and $b$-axis, respectively. Equation (2) above describes the rotation of the polarization between the $b$-axis ($\beta$=0) and the $a$-axis ($\beta$=90$^o$), as observed in our experiments. Focussing on the angular dependence ($\beta$,$\alpha$), it can be written as
\begin{equation}
\roarrow P = P_{0}(T)(sin(\beta)\roarrow a+cos(\beta)sin(\alpha)\roarrow b)
\end{equation}
The rotation angle of the helical plane $\beta$ is related to the measured polarization angle $\Theta_{b}$ through tan($\beta$) = tan($\Theta_{b}$)sin($\alpha$). Equation (3) should qualitatively describe the data shown in Fig. 2, however, for a comparison some simplifying assumptions have to be made. The angle $\alpha$ is chosen as 16$^o$ (consistent with neutron scattering results\cite{Song:10,Ye:11}) and no change of this angle with magnetic field is considered. Furthermore, we assume that the product $m_{b}m_{easy}$ does not change in external fields. As to the temperature dependence of $P_{0}(T)$ in equation (3) we assume the mean field function $P_{0}(T)=P_{0}(0)\sqrt{1-T/T_{C}}$, with $P_{0}(0)= 90 \mu C/m^2$. The magnitude of the polarization vector can now be expressed as

\begin{equation}
\left|P(T)\right| = P_{0}(T)\sqrt{(sin^2(\beta)+cos^2(\beta)sin^2(\alpha))}
\end{equation}

With the experimental data for the temperature dependence of $\Theta_{b}$ (Fig. 2) as input to equation (4) the magnitude of the polarization is calculated and shown in the inset to Fig. 2 (labeled "Simulation"). The temperature dependence of the measured $\left|P(T)\right|$ and the characteristic features including the maximum of $\left|P(T)\right|$ below $T_{C}$ are in fact very well reproduced. A more quantitative evaluation does require additional input from high-field neutron scattering experiments such as the possible field and temperature dependence of the parameters defining the helical spin structure ($\alpha, m_{b}, m_{easy}$, etc.) and a low-temperature correction to the mean field formula for $P_{0}(T)$. However, the good agreement of the experimental data with the simple formula for the polarization (equation(4)) prove that the orientation of the spin helical plane is the major parameter determining the magnitude and orientation of $\roarrow P(T)$.

%\section{Summary and conclusions}
For Mn$_{0.95}$Co$_{0.05}$WO$_{4}$, the AF1 phase is no longer stable in the field range of up to 7 Tesla. Instead, the continuous rotation of the ferroelectric polarization toward the $a$-axis has been observed above 2 Tesla. This magnetic-field effect on the polarization direction results in a new phenomenon, the rotation of the polarization upon decreasing temperature in a constant magnetic field and a maximum of the polarization magnitude below $T_{C}$. This is qualitatively understood by the rotation of the spin helix about the easy axis of magnetization. The formulas have been derived to take into account the additional rotation angle and the calculated polarization function of temperature is in very good agreement with the data. The complex dependence of the ferroelectric polarization on magnetic field and temperature is different from the field-induced polarization flop observed in MnWO$_{4}$. In the undoped MnWO$_{4}$, an $a$-axis polarization was detected only at much higher fields, H$_{b} >$10 Tesla, and in a very narrow temperature range between 10 K and 11 K but the associated magnetic structure was not explored.\cite{Taniguchi:06} It appears reasonable to assume that the magnetic structures leading to an $a$-axis ferroelectric state in MnWO$_{4}$ and Mn$_{0.95}$Co$_{0.05}$WO$_{4}$ are very similar and that the largely extended stability of this phase in the Co-doped compound is directly related to the Co substitution.

%\label{}
%\subsection{}
%\subsubsection{}
% If in two-column mode, this environment will change to single-column format so that long equations can be displayed.
%\begin{widetext}
%$$\mbox{put long equation here}$$
%\end{widetext}
% Figures should be put into the text as floats.
% Use the graphics or graphicx packages (distributed with LaTeX2e).
% Put the label that you will use with \ref{} command in the braces of the \label{} command.
% \begin{table}
% \end{tabular}
% \end{table}
\begin{acknowledgments}
This work is supported in part by the T.L.L. Temple Foundation, the John J. and Rebecca Moores Endowment, the State of Texas through TCSUH, the U.S. Air Force Office of Scientific Research, and at LBNL through the U.S. Department of Energy.
\end{acknowledgments}

\bibliographystyle{phpf}

%\bibliography{56MMM}
\end{document}